\documentstyle[aps,prb,epsf,eqsecnum]{revtex}
\begin{document}
\draft

% comment out the following line for single column output
%\twocolumn[\hsize\textwidth\columnwidth\hsize\csname @twocolumnfalse\endcsname

\title{Logarithmic Corrections in Quantum Impurity Problems}

\author{
Ian Affleck$^{1,2}$ and Shaojin Qin$^{1,3}$
}
\address{
$^1$Department of Physics and Astronomy, 
The University of British Columbia, Vancouver, B.C., V6T 1Z1, Canada \\
$^2$Canadian Institute for Advanced Research, 
The University of British Columbia, Vancouver, B.C., V6T 1Z1, Canada \\
$^3$ Institute of Theoretical Physics, 
P.O. Box 2735, Beijing 10080, People's Republic of China
}

\date{\today}

\maketitle

\begin{abstract}
The effect of a {\it bulk} marginal operator on {\it boundary} 
critical phenomena in two space-time dimensions is considered.  
The particular case of an open S=1/2 antiferromagnetic Heisenberg 
chain, corresponding to a Wess-Zumino-Witten non-linear $\sigma$ model, 
is solved.  In this case, the needed renormalization group coefficient 
is associated with a novel operator product expansion in which 
{\it three} operators approach the same point.  Resulting logarithmic 
corrections occurring in finite size calculations and nuclear magnetic 
resonance experiments are discussed.  
\end{abstract}


% comment out the following line for single column output
% ]
\section{Introduction}
Marginally irrelevant operators in two-dimensional conformal field \cite{Cardy}
theory lead to logarithmic corrections to scaling behaviour.  Because 
the corresponding coupling constant, $g(l)$, renormalizes to 0 very 
slowly, as $1/\ln l$ where $l$ is a characteristic length or energy 
scale, logarithmic corrections occur to virtually {\it all} quantities 
which can be measured experimentally or simulated numerically.  These 
create grave difficulties in obtaining agreement between analytical 
theory and numerical simulations or experiment.  The particular case 
of the S=1/2 Heisenberg antiferromagnetic chain has been discussed 
extensively.  The correlation function, initially predicted to decay 
as $1/r$, instead decays as\cite{Affleck1,Singh} $(\ln r)^{1/2}/r$.  To make matters worse, 
the corrections to this result are only suppressed by additional powers 
of $1/\ln r$ and are highly sensitive to finite-size effects.  
Similarly, the energy gap between the ground state and the first 
excited (triplet) state behaves as \cite{Cardy}:
\begin{equation}
\Delta E = {2\pi v\over l}
	\left[{1\over 2} -{\pi g(l)\over \sqrt{3}}\right] \,
\label{gcorr}
\end{equation}
where $v$ is the spin-wave velocity.  At very long lengths,
\begin{equation}
g(l)\to {\sqrt{3}\over 4\pi \ln l},
\label{gas}
\end{equation}
giving an additive logarithmic correction to the finite-size energy 
gap.  It is generally very hard to actually observe this logarithmic 
behaviour unless chains of length several thousand can be studied.  
Fortunately, this is possible for Bethe ansatz integrable models like 
the S=1/2 Heisenberg chain.  For shorter chains it is generally better 
to regard $g(l)$ as a free parameter.  Because Eq. (\ref{gcorr}) has 
been generalized to many other energy levels, all of which receive 
corrections linear in $g$, this still has considerable predictive 
power.  Indeed, fitting to expressions of this sort, which effectively 
subtracts off the leading logarithmic correction to scaling, provides 
a practical method for numerically determining the universality class 
of a Hamiltonian. \cite{Affleck1}  There are analogous finite temperature corrections 
(for infinite length systems).  These include an additive $1/\ln T$ 
correction to the susceptibility \cite{Eggert1} and a multiplicative $(\ln T)^{1/2}$ 
correction to the spin relaxation rate, $1/T_1$. \cite{Sachdev}  Although experimental 
data has been fit to these forms with apparent success \cite{Motoyama,Takigawa1} this is a very 
difficult exercise due to the slow variation and the presence of 
various other types of corrections in any real material.

Another subject of current experimental and theoretical interest is the 
general area of quantum impurity problems (QIP's).  In the context of 
S=1/2 Heisenberg antiferromagnetic chains, such a problem occurs for a 
semi-infinite chain, associated with the dynamics at the chain end.\cite{Eggert2}  
This model can be realized experimentally by dilute substitution of the 
magnetic ion by a non-magnetic one, $eg.$ Zn substitution for Cu.  
Related QIP's involve the Kondo problem and tunneling through a single 
impurity in a quantum wire (or quantum Hall effect edge states).  The 
general renormalization group (RG) treatment of these problems gives 
fixed points corresponding to conformally invariant boundary conditions 
imposed on a given bulk conformal field theory.\cite{Affleck2}  In general, in such a 
theory, the effective Hamiltonian contains both bulk and boundary 
operators.  The bulk terms contain integrals over the half-line whereas 
the boundary terms occur at the impurity location, $x=0$.  Bulk 
behaviour is unaffected by boundary dynamics, although it must be 
appreciated that the decay of Green's functions away from the boundary 
is itself part of the boundary critical phenomena.  Time correlation 
functions at the boundary also involve exponents which characterize the 
boundary condition, as does the finite size spectrum with non-periodic 
boundary conditions.  Boundary interactions cannot affect the 
renormalization of bulk coupling constants.  In most treatments of these 
problems so far, any renormalization of boundary interactions by bulk 
interactions has also been ignored.  The justification for this is that 
the bulk system has been assumed to be at a bulk RG fixed point.  Any 
bulk operators present, apart from the fixed point Hamiltonian itself, 
are irrelevant and can be ignored at low energies.  Thus, at least in 
principle, crossover between boundary fixed points can be treated 
independently of bulk renormalization.  Strictly speaking, this is 
only justified when the energy scales associated with the boundary 
renormalization are much smaller than those associated with the bulk 
irrelevant couplings.  This approximation is particularly bad when there 
are marginally irrelevant bulk interactions present since they 
renormalize to 0 logarithmically slowly.  When marginally irrelevant 
bulk operators are present we should expect logarithmic corrections to 
exponents and finite-size scaling.  However, the detailed form of these 
corrections is characteristic of the boundary condition and is not 
simply related to the log corrections in the bulk theory, nor to the 
finite-size scaling with periodic boundary conditions.  

It is the purpose of this note to consider these logarithmic 
corrections to boundary critical exponents and finite-size scaling 
with non-periodic boundary conditions arising from a marginal bulk 
operator.  In the next section we calculate general formulas for
the logarithmic corrections using conformal field theory.  In Section 3
we compare these formulas to results on the finite size spectrum for
an S=1/2 chain with open boundary conditions, for lengths up to 2,000, obtained
from the Bethe ansatz.  In the final section we comment on the log corrections
to correlation functions and the NMR relaxation rate.
\section{Conformal Field Theory Results}
As observed by Cardy,\cite{Cardy} such logarithmic corrections generally result 
from corrections to the anomalous dimensions of the various operators, 
$\phi_n$, which are linear in $g$.  The associated coupling constants, 
$u_n$, obey an renormalization group (RG) equation:
\begin{equation}
du_n/d\ln L = [2-\gamma_n ]g,
\end{equation}
where:
\begin{equation}
\gamma_n=x_n+2\pi b_ng + \ldots 
\label{anom}
\end{equation}
Here the $\ldots$ represents terms of higher order in $g$ or in other 
irrelevant operators.  Taking into account the fact that $g$ itself 
also renormalizes leads to predictions of the various logarithmic 
corrections.  In particular, the finite-size states are in one to one 
correspondence with the operators and their energy gaps are 
proportional to $\gamma_n$.  The coefficients, $b_n$ can be 
conveniently determined from the operator product expansion (OPE) of 
the operator $\phi_n$ with the marginal operator, $\phi$:
\begin{equation}
\phi (z)\phi_n(z')\to {-b_n \phi_n(z)\over |z-z'|^2}.
\label{bndef}
\end{equation}
A calculation of any Green's function involving $\phi_n$, to the first 
order in $g$, encounters a logarithmic ultraviolet divergence upon 
integrating $z$ near $z'$.  This implies the correction to the 
anomalous dimension in Eq. (\ref{anom}).  The finite size spectrum is 
given by:
\begin{equation}
E_n-E_0\approx {2\pi v\over l}[x_n+2\pi b_ng(l)].
\label{fss}
\end{equation}
The scaling dimension $x_n$ of the corresponding operator is simply 
corrected by the anomalous dimension term of first order in $g$ which 
is replaced by the effective coupling at scale $l$.  

In the presence of a boundary, this calculation takes a rather 
unfamiliar turn because the local marginal bulk operator in 
general becomes {\it bilocal} in the presence of a boundary condition.  
This follows from Cardy's general approach to conformally invariant 
boundary conditions, which are always assumed to obey:
\begin{equation} 
T_L(t,0)=T_R(t,0),  
\end{equation}
where $T_{L,R}$ are the left and right moving terms in the energy 
momentum tensor.  Since $T_{L,R}$ is a function of $(t-x)$ [$(t+x)$] 
only, it then follows that we may regard the right-movers on the 
original physical space, $x>0$ as the continuation of the left-movers 
to the negative axis,
\begin{equation}
T_R(x)=T_L(-x), \ \  (x>0).
\label{T}
\end{equation}
This observation allows the Hamiltonian to be written in terms of left 
movers only, but defined on the entire real line.  In particular, it 
implies that a generic local bulk operator, which can be factorized 
into its left-moving and right-moving parts, $O_L^1(t-x)$ and 
$O_R^2(t+x)$ respectively, becomes {\it bilocal}:
\begin{equation}
O(t,x)=O_L^1(t,x)O_R^2(t,x)\to O_L^1(t,x)O_L^2(t,-x).
\label{non-local}
\end{equation}
In particular, the bulk marginal operator becomes bilocal, introducing 
a novel complication in calculating its effects perturbatively.  
Boundary operators are also drawn from the left-moving sector only.  
We may determine the correction to the anomalous dimension of an 
arbitrary boundary operator, $\Phi$, to first order in $g$, from the 
3-point function $<\Phi O \Phi^\dagger >$, where $O$ is the marginal 
bulk operator.  However, this 3-point function must be calculated in 
the presence of the boundary condition, upon which it depends.  
Furthermore, we see from Eq. (\ref{non-local}) that this 3-point 
function effectively becomes a 4-point function of left-moving 
operators.  

In general, some data about the boundary condition will be needed to 
calculate this 4-point function.  As will be seen from the example 
considered below, it is sufficient to know the OPE of the chiral part 
of the bulk marginal operator, $O(z)$ with general boundary operators.  
We consider a special case here, of some importance,  for which this 
OPE can be readily calculated. 

Let us consider the problem of an S=1/2 Heisenberg chain with an open 
boundary condition. The bosonized form of this model, applicable at low 
energies, is the $k=1$ Wess-Zumino-Witten non-linear $\sigma$ model.  
The marginal operator is quadratic in the chiral spin densities, 
$\vec J_{L,R}$, and is written:
\begin{equation}
H=H_0-g(8\pi^2/\sqrt{3})\vec J_L\cdot \vec J_R .
\label{Hnorm}
\end{equation}
Here the spin densities are normalized as:
\begin{equation}
<J_L^a(z)J_L^b(0)>={\delta^{ab}\over 8\pi^2z^2}.
\label{Jnorm}
\end{equation}
The factor of $8\pi^2/\sqrt{3}$ is inserted so that the operator 
multiplied by $g$ in the Hamiltonian has a unit-normalized 2-point 
function, following the convention of Cardy.\cite{Cardy}  The needed OPE's in this 
case follow from the basic one of the WZW model:
\begin{equation}
\vec J_L^a(z)\phi(z') = {\vec S_L \phi (z)\over 2\pi (z-z')}+ \ldots 
\end{equation}
Here the general Virasoro primary operator, $\phi$, may transform under 
an arbitrary representation of $SU(2)_L\times SU(2)_R$.  The finite 
dimensional matrices, $S_L^a$ are simply the representation of SU(2) 
under which $\phi$ transforms.  Explicitly, if $\phi$ transforms under 
the irreducible spin $S$ representation then there will be a multiplet 
of $(2S+1)$ operators, $\phi_A$ and
\begin{equation} 
(\vec S_L\phi )^A\equiv \sum_B\vec S_L^{AB}\phi^B.
\end{equation}
Thus,
\begin{equation}
\vec J_L(z)\cdot \vec J_R(z)\phi (z') = 
	{\vec S_L\cdot \vec S_R\over 4\pi^2|z-z'|^2}\phi (z)+\ldots 
\label{OPE}\end{equation}
and we see that the coefficient $b_n$ define in Eq. (\ref{bndef}) takes 
the value:
\begin{equation}
b_n=-{2\vec S_L\cdot \vec S_R\over \sqrt{3}}.
\label{bn}
\end{equation}

Now consider a semi-infinite chain with a free boundary condition at 
the origin.  This boundary condition was treated using bosonization in 
Ref. [\onlinecite{Eggert2}].  The boundary condition not only is 
consistent with Eq. (\ref{T}) but also with its Kac-Moody 
generalization:
\begin{equation}
\vec J_R(x)=\vec J_L(-x).
\label{J} 
\end{equation}
Eq. (\ref{T}) and (\ref{J}) imply that the correlation function of 
$T(t,x)$ and $\vec J(t,x)$ are simply the chiral correlation functions 
in the free WZW model.  The boundary has no affect on them apart from 
identifying left with right.  The operator content and finite size 
spectrum are drawn from conformal towers of the left-moving Kac-Moody 
algebra only.  It was shown in [\onlinecite{Eggert2}] that, for an even 
length chain, the spectrum consists of the identity conformal tower 
only.  In particular, the spin operator at the boundary becomes the 
(left-moving) spin density, $\vec J_L(t,0)$ of the WZW model with 
correlation function, $\propto 1/t^2$.  The finite-size spectrum, with 
free boundary conditions at both ends of a chain of length $l$ is given 
by:
\begin{equation}
E-E_0= {\pi v\over l}x_L,
\label{fsso}
\end{equation}
where $x_L$ is the scaling dimension of the corresponding (left-moving) 
operator.  This differs from the formula with periodic boundary 
conditions, Eq. (\ref{fss}), by the replacement of $l$ by $2l$ 
corresponding to doubling the system size due to the identification of 
Eq. (\ref{T}) at $x=0$ and $x=l$ and to the replacement of the scaling 
dimension $x=x_L+x_R$ by a left-moving scaling dimension $x_L$ only.  

Let us now consider the logarithmic corrections to this formula for 
open boundary conditions.  The problem again reduces to finding the 
correction to the anomalous dimension of a given operator, $\phi_n$,
to first order in $g$ where $\phi_n$ is now a {\it boundary} operator.  
We might again attempt to obtain this from an OPE but we must deal with 
the fact that the marginal operator is now {\it bilocal}.  The obvious 
generalization of Eq. (\ref{OPE}) is:
\begin{equation}
\vec J_L(z)\cdot \vec J_L(z^*)\phi(0) \approx 
	{\vec S_L\cdot \vec S_L\over |2\pi z|^2}\phi .
\label{3OPE}
\end{equation}
However, this is not a conventional OPE because we are bringing 
{\it three} operators to the same point rather than just two.  However, 
as will be argued below, it is nonetheless correct.  It is now easy to 
calculate the correction to the anomalous dimension.  A logarithmic 
ultraviolet divergence is again encountered for any Green's function 
involving the operator $\phi$, to first order in $g$ from integrating 
over $z$ near $0$.  However this is reduced by a factor of two relative 
to the bulk case because $z$ is only integrated over the half-plane 
rather than the entire plane. A convenient ultraviolet cut-off is given 
by restricting the z integral to $|z|>a$.  In the bulk case the 
excluded region is a circle of radius $a$ around the origin but in the 
boundary case it is only a semi-circle of radius $a$.  Our conclusion 
is thus that the anomalous dimension of a boundary operator is given by 
Eq. (\ref{anom}) but with the coefficient $b_n$ now given by:
\begin{equation}
b_n=-{\vec S_L\cdot \vec S_L\over \sqrt{3}}.
\label{bnb}
\end{equation}
This differs from the bulk formula Eq. (\ref{bn}) only by the 
identification of $\vec S_R$ with $\vec S_L$ and by the extra factor of 
$1/2$ arising from the different integration region.

To complete our derivation we just need to justify the rather 
unorthodox 3-operator OPE occurring in Eq. (\ref{3OPE}).  The validity 
of this formula can be understood by considering the more general 
connected 4-point function:
\begin{equation}
G^{AB} = <\vec J_L(z_1) \cdot \vec J_L(z_2)
		\phi^A(0)\phi^B(\tau )>_{\hbox{connected}}.
\end{equation}
We normalize the Virasoro primary boundary operator, $\phi^A$ so that 
its two point function is given by:
\begin{equation}
<\phi^A(0)\phi^B(\tau )> = {\delta^{AB}\over (-\tau )^{2x_L}}.
\end{equation}
We wish to consider the short-distance singularity in $G^{AB}$ when 
$z_1, z_2\to 0$.  We claim that this is given by:
\begin{equation}
G^{AB} \to {\vec S_L \cdot \vec S_L \over 4\pi^2z_1z_2}
			{\delta^{AB}\over (-\tau)^{2x}}.
\label{sing}
\end{equation}
The correctness of this result can be seen by considering the three 
different limits $|z_1|<<|z_2|, |z_1-z_2|$ and $|z_2|<<|z_1|,|z_1-z_2|$ 
and $|z_1-z_2|<<|z_1|,|z_2|$.  In the first case, we can obtain the 
leading singularity by using the OPE of $\vec J_L(z_1)$ with 
$\phi^A(0)$ and then the OPE of the result with $\vec J_L(z_2)$.  This 
gives Eq. (\ref{sing}).  The same singularity is obtained in the second 
case.  In the third case there should be no singularity of the form 
$1/(z_1-z_2)$ because there is no singular term in the OPE 
$\vec J_L(z_1)\cdot \vec J_L(z_2)$ apart from the trivial one which 
doesn't contribute to the connected Green's function.  These 
considerations uniquely fix all singularities in $G^{AB}$ at 
$z_1, z_2\to 0$.  Note that the crucial property of the boundary 
condition that is being used is that the OPE of the spin density 
operators $J_L^z$ with arbitrary (Virasoro primary) boundary operators 
has the same form as in the bulk.  Now letting $z_1=z$, $z_2=z^*$ gives 
Eq. (\ref{3OPE}).

In the particular case where the $\phi_A$ are the spin-density 
operators, $J_L^a$, we have calculated $G^{ab}$ exactly and verified 
the form of the singularity.  In this case we find:
\begin{equation}
<\vec J_L(z_1)\cdot \vec J_L(z_2)J_L^a(0)J_L^b(\tau )> = 
	{\delta^{ab}\over (2\pi )^4z_1z_2(\tau-z_1)(\tau -z_2)}.
\label{J4pt}
\end{equation}
We see from Eq. (\ref{Jnorm}) that the unit normalized operator is 
$\phi^a=2\pi\sqrt{2}J^a$. Also using the fact that $\vec J$ has $S_L=1$ 
and therefore $\vec S_L\cdot \vec S_L=S_L(S_L+1)=2$, we see that, in 
the limit $z_i\to 0$, Eq. (\ref{J4pt}) agrees with Eq. (\ref{sing}).

Now let us consider the finite size spectrum, examining the lowest 
energy excited state of spin $S$. This is given by Eq. (\ref{fsso}) 
except that the dimension of the (left-moving) field, $x_L$ must be 
replaced by the anomalous dimension, $\gamma_n$.  This is given by 
Eq. (\ref{anom}) with $b_n$ now given by Eq. (\ref{bnb}).  Thus we 
obtain:
\begin{equation}
E_S^{\hbox{open}}-E_0^{\hbox{open}} \approx 
	{\pi v\over l}\left[ S^2-{2\pi S(S+1)g(l)\over \sqrt{3}}\right].
\label{oglz}
\end{equation}
For exponentially long chains we may use the asymptotic form of $g(l)$: 
$g(l)\to \sqrt{3}/(4\pi \ln l )$, giving:
\begin{equation}
E_S^{\hbox{open}}-E_0^{\hbox{open}} \approx 
	{\pi v\over l}\left[ S^2-{S(S+1)\over 2\ln l}\right].
\label{ofss}
\end{equation}
It is interesting to compare Eq. (\ref{ofss}) to the corresponding 
result for periodic boundary conditions.  \cite{Affleck1}  If we again 
consider the lowest energy state of given spin $S$, this has 
$S_L=S_R=S/2$ and hence:
\begin{equation}
E_S^{\hbox{per}}-E_0^{\hbox{per}} \approx 
	{2\pi v\over l}\left[{S^2\over 2}-{S^2\over 4\ln l}\right] .
\end{equation}
The $1/l$ terms are the same for open and periodic boundary conditions 
but the $1/\ln l$ terms are not.  (We note that the logarithmic 
corrections for open boundary conditions were assumed to be same as the 
ones for periodic boundary conditions in [\onlinecite{Ng})].

We have also calculated the logarithmic correction to the ground state 
energy for open boundary conditions.  Ignoring the irrelevant operator, 
the ground state energy for any one dimensional Hamiltonian which 
renormalizes to a conformal field theory, defined on an interval of 
length $l$ with generic boundary conditions at $0$ and $l$ consistent 
with Eq. (\ref{T}) is:
\begin{equation}
E_0(l)=e_0l+e_1-(\pi v/24l)c,
\label{E0}
\end{equation}
where $c$ is the central charge.\cite{Blote}  Note that the coefficient 
of $1/l$ is 1/4 times the value for periodic boundary conditions.  Also 
note that an additional non-universal surface energy, $e_1$ appears 
when the boundary conditions are non-periodic.  Logarithmic 
corrections to this formula can be calculated by doing perturbation 
theory in the marginally irrelevant coupling constant, $g(l)$ just as
in the periodic case.  One finds that the correction of $O(1/l)$ is 
universal.  This must be separated from various non-universal 
corrections to $e_0$ and $e_1$.  Once the correction to the $1/l$ term 
of leading order in $g$ is calculated, $g$ may be replaced by the 
effective coupling constant at scale $l$, $g(l)$, resulting in 
logarithmic corrections.  In the case of periodic boundary conditions, 
this leading correction was found to be $O(g^3)$.  By contrast, in the 
case of open boundary conditions we find that it is $O(g^2)$.

Let us first consider the correction of $O(g)$.  From Eq. (\ref{Hnorm}) 
this gives a correction to the ground state energy:
\begin{equation}
\delta E_0^{(1)} = 
{-8\pi^2g\over \sqrt{3}}\int_0^ldx<\vec J_L(x)\cdot \vec J_L(-x)>.
\label{E1int}
\end{equation}
This Green's function is given in Eq. (\ref{Jnorm}) for the case 
$l\to \infty$.  We may obtain the Green's function for finite length by 
a conformal transformation:
\begin{equation} 
(\tau '+ix')=e^{(\pi /l)(\tau + ix)},
\label{CT}
\end{equation} 
giving:
\begin{equation}
<J_L^a(x)J_L^b(-x)> = 
{\delta^{ab}\over 8\pi^2\left[{l\over \pi}\sin {\pi x\over l}\right]^2}.
\end{equation}
The integral of Eq. (\ref{E1int}) is ultraviolet divergent both at 
$x=0$ and $x=l$.  We may insert an ultraviolet cut off on the 
integration region, $x>a$, $l-x>a$ where $a$ is of order the lattice 
spacing.  This gives the ground state energy correction of first order 
in $g$:
\begin{equation}
\delta E_0^{(1)}=-2g\sqrt{3}{\pi \over l}\cot {\pi a\over l}.
\end{equation}
Now Taylor expanding in powers of $a/l$, we see that we obtain a cut 
off dependent contribution to the non-universal surface energy, $e_1$ 
in Eq. (\ref{E0}) together with corrections of $O(1/l^2)$:
\begin{equation}
\delta E_0^{(1)}\approx {-2g\sqrt{3}\over a}+ O(a/l^2).
\end{equation}
Importantly, there is no term of $O(1/l)$.

We now push this calculation to second order in $g$.  This term is 
given by:
\begin{equation}
\delta E_0^{(2)} = -{1\over 2} \left[{8\pi^2\over \sqrt{3}}g\right]^2
	\int_{-\infty}^\infty d\tau \int_0^l dx_1\int_0^l dx_2 
		<\vec J_L(\tau ,x_1)\cdot \vec J_L(\tau ,-x_1)
			\vec J_L(0,x_2)\cdot \vec J_L(0,-x_2)>.
\end{equation}
Let us first evaluate this expression in the limit $l\to \infty$.  
Using Eq. (\ref{J4pt}), we obtain:
\begin{equation}
\delta E_0^{(2)} \to -\left[{8\pi^2\over \sqrt{3}}g\right]^2
	{3\over (2\pi )^4}\int_{-\infty}^\infty d\tau \int_0^\infty 
	{dx_1dx_2\over [\tau^2+(x_1-x_2)^2][\tau^2+(x_1+x_2)^2]}.
\end{equation}
Note that we have inserted a factor of $2$ here because equal 
contributions arise from $x$ near $0$ and $x$ near $l$.  The 
$x$-integrals can be done exactly and are ultraviolet finite, for 
non-zero $\tau$:
\begin{equation}
\delta E_0^{(2)} \to 
	-{g^2\pi^2\over 2}\int_{-\infty}^\infty {d\tau \over \tau^2}.
\end{equation}
The $\tau$ integral is ultraviolet divergent.  We cut off the integral, 
$|\tau |>\tau_0$, where $\tau_0$ is of $O(a/v)$.  This gives the second 
order ground state energy correction:
\begin{equation}
\delta E_0^{(2)}\to -{g^2\pi^2\over \tau_0},
\label{E02inf}
\end{equation}
another cut-off dependent contribution to the surface energy, $e_1$ in 
Eq. (\ref{E0}).  To obtain $\delta E_0^{(2)}$ at finite $l$, we again 
use the conformal transformation of Eq. (\ref{CT}) to obtain:
\begin{equation}
\delta E_0^{(2)} = -2g^2\int_{-\infty}^\infty d\tau \int_0^l
	{dx_1dx_2(\pi /2l)^4 \over 
		|\sin (\pi /2l)[(x_1-x_2)^2+i\tau ]|^2
		|\sin (\pi /2l)[(x_1+x_2)^2+i\tau ]|^2}.
\end{equation}
The $x_i$ integrals are again finite for non-zero $\tau$.  We again cut 
off the $\tau$ integral at $|\tau |>\tau_0$.  Noting that the integrand 
is symmetric under $x_1\to -x_1$ or $x_2\to -x_2$ and also 
$\tau\to -\tau$, it is convenient to extend the $x_i$ integrals from 
$-l$ to $l$ and reduce the $\tau$ integral from $\tau_0$ to $\infty$.
This introduces a net factor of $1/2$.  It is now convenient to change 
variables to:
\begin{equation} 
z_j\equiv e^{i\pi x_j/l},
\end{equation}
and
\begin{equation}
u\equiv\pi \tau /l.
\end{equation}
The complex variables, $z_j$ are integrated around the unit circle.  
This expression now becomes:
\begin{equation}
\delta E_0^{(2)} = -{g^2\pi \over l} \int_{u_0}^{\infty}du 
	\int_Cdz_1\int_Cdz_2 {z_1 \over 
		z_2(z_1-z_2e^u)(z_1-z_2e^{-u})
		(z_1-z_2^{-1}e^u)(z_1-z_2^{-1}e^{-u})},
\end{equation}
where $u_0\equiv \pi \tau_o/l$ and $C$ denotes the unit circle 
integration contour.  The $z_1$ integral can now be done by the 
standard contour integration method, with contributions from the poles 
at $z_2^{\pm 1}e^{-u}$.  The result is:
\begin{equation}
\delta E_0^{(2)} = -2\pi ig^2{\pi \over l}
	\int_{u_0}^\infty du \coth u \int_Cdz_2
		{z_2\over (z_2^2-e^{2u})(z_2^2-e^{-2u})}.
\end{equation}
Also doing the $z_2$ integral by contour methods gives:
\begin{equation}
\delta E_0^{(2)} = -\pi^2g^2(\pi /l)\int_{u_0}^\infty 
	{du \over \sinh^2(u)}.
\end{equation}
Finally, performing this elementary integration gives:
\begin{equation}
\delta E_0^{(2)}=-2\pi^2g^2(\pi /l){1\over e^{2u_0}-1}\approx 
	-{\pi^2g^2\over \tau_0}+\pi^2g^2{\pi \over l} + O(\tau_0/l^2).
\end{equation}
We have recovered the same cut off dependent to the surface energy, 
$e_1$ as in Eq. (\ref{E02inf}).  More importantly, we have also 
obtained a term of $O(1/l)$ {\it which is cut off independent} and 
therefore is expected to be universal.  Thus, we obtain the log 
correction to the ground state energy with open boundary conditions:
\begin{equation}
E_0^{\hbox{open}}(l)\approx e_0l+e_1-{\pi v\over 24 l}[1-24\pi^2g(l)^2].
\label{oglgs}
\end{equation}
For exponentially large $l$ we may use Eq. (\ref{gas}) to write:
\begin{equation}
E_0^{\hbox{open}}(l) \approx e_0l+e_1
	-{\pi v\over 24 l}\left[ 1-{9/2\over (\ln l)^2}\right] .
\end{equation}
As usual, the corrections to these formulas are only down by additional 
powers of $g(l)$, that is $1/\ln l$.  The corresponding formula for 
periodic boundary conditions is:
\begin{equation}
E_0^{\hbox{per}}(l) 
\approx e_0l-{\pi v\over 6 l}[1+(2\pi )^3g(l)^3/\sqrt{3}]
\approx e_0l-{\pi v\over 6 l}\left[ 1+{3/8\over (\ln l)^3}\right] .
\end{equation}
Among other differences, note that the log corrections {\it decrease} 
the apparent value of $c$ for open boundary conditions but 
{\it increase} it for periodic boundary conditions.  The fact
that the correction to $c$ goes like $1/(\ln l)^2$ was obtained
from the Bethe ansatz in Ref. [\onlinecite{Hamer}], although the coefficient
was not obtained.

\section{Numerical Results on Finite Size Spectrum}
One application of the above results for boundary critical phenomena is 
to the numerical study of finite size 
scaling.  We extract estimates of $g(l)$ defined in Eq.(\ref{oglz}) and 
Eq.(\ref{oglgs}) from the energies of finite size spin 1/2 
antiferromagnetic Heisenberg open chains.  The Hamiltonian is
\begin{equation}
H = \sum_{i=1}^{l-1}{\bf S}_i\cdot{\bf S}_{i+1}, 
\label{hms05}
\end{equation}
where the ${\bf S}_i$'s are spin 1/2 operators.  The Bethe ansatz 
equations\cite{Gaudin} for the Hamiltonian of Eq.~(\ref{hms05}) are
\begin{equation}
\left(\frac{\Lambda_k+i/2}{\Lambda_k-i/2}\right)^{2l} =
\prod_{j\not =k}^M 
\frac{\Lambda_k-\Lambda_j+i}{\Lambda_k-\Lambda_j-i}\;\;
\frac{\Lambda_k+\Lambda_j+i}{\Lambda_k+\Lambda_j-i},
\label{bae}
\end{equation}
where $l$ is the number of sites in the open chain.  The roots can 
be numerically calculated.\cite{Sorensen}  The number of roots, $M$ 
determines the total $S^z$ component through the relation 
$S^z=L/2-M$.  In terms of the solutions, $\Lambda_k$, to the Bethe 
ansatz equations, Eq.~(\ref{bae}), the energy is 
given by 
\begin{equation}
E={\frac{l-1}{4}}
-{\frac 1 2}\sum_{k=1}^M{\frac 1 {\Lambda_k^2+1/4}}. 
\end{equation} 

The surface energy $e_1=(\pi-1-2\ln 2)/4$ for the two ends of open 
chain can be exactly obtained.  The rapidities 
$\{\Lambda_k,k=1,l/2\}$ solving the Bethe ansatz equations 
Eq.~(\ref{bae}) for the ground state of the open chain are all bigger than 
zero. Let's order $\Lambda_k$ so that: 
$\Lambda_1>\Lambda_2>\ldots>\Lambda_{l/2-1}>\Lambda_{l/2}>0$.  We can 
construct another set of rapidities $\{\Lambda '_k,k=1,l\}$:
\begin{equation}
\Lambda '_{j}=-\Lambda_j,\;\;\;\; 
\Lambda '_{l+1-j}=\Lambda_j,\;\;\;\; 
{\rm for} \;\;\;\; j=1 \;\;\;\; {\rm to} \;\;\;\; l/2,
\end{equation}
which solves exactly the following Bethe ansatz equations for a periodic 
chain of length $2l+1$:
\begin{equation}
\left(\frac{\Lambda '_k+i/2}{\Lambda '_k-i/2}\right)^{2l+1} =
\prod_{j\not =k}^{l}
\frac{\Lambda '_k-\Lambda '_j+i}{\Lambda '_k-\Lambda '_j-i}.
\end{equation}
The energy for the periodic chain is given by 
\begin{equation}
E'={\frac{2l+1}{4}}
-{\frac 1 2}\sum_{k=1}^{l}{\frac 1 {{\Lambda '}_k^2+1/4}}. 
\end{equation} 
In the logarithmic form of the Bethe ansatz equations for the periodic chain, 
the set $\{\Lambda '_k,k=1,l\}$ corresponds to a set of integers
$\{I '_k,k=1,l\}$:
\begin{eqnarray}
I '_{j}=j-(l/2+1),\;\;\;\; 
I '_{l+1-j}= j,\;\;\;\; 
{\rm for} \;\;\;\; j=1 \;\;\;\; {\rm to} \;\;\;\; l/2, \nonumber \\
i.e.,\;\;\;\;
-l/2,\; -l/2+1,\; \ldots -2,\; -1,\;
\;\;1,\; 2,\; \ldots l/2-1,\; l/2.
\end{eqnarray}
The groundstate of the open chain corresponds to an excited state
of the periodic chain 
 with a hole exactly at $I_i=0$ in the
connected integer set $\{I_i,i=1,l+1\}$ for the ground state of the periodic 
chain.\cite{Faddeev}  While the groundstates of the periodic chain of
odd length have spin $S^z=\pm 1/2$ and momentum $\pm \pi /2$, the state
of the odd length periodic chain corresponding to the groundstate of
the even length open chain has $S^z=-1/2$ and momentum 0.  From its
momentum, and from the fact that it has one hole we expect
its energy, in the large $l$ limit, to be the groundstate energy
of the periodic chain plus the excitation energy for a magnon of
momentum $\pi /2$:
$E'=e_0 (2l+1) + v_s \sin(k)$, $v_s=\pi/2$, and $e_0=1/4-\ln 2$ for the 
spin 1/2 chain.  Comparing the expressions for $E$ and $E'$, we can 
eliminate the summation over $\Lambda$ and obtain $E=e_0l+e_1$ for the 
ground state of the open chain with $e_1=(\pi-1-2\ln 2)/4$. We have
checked this result from our numerical solution of the Bethe ansatz
equations for finite $l$ obtaining agreement to at least 6 decimal places.  
This result was derived earlier from the Bethe ansatz equations by
a somewhat different method.\cite{Hamer,Alcarez,Batchelor}

\begin{figure}[ht]
\epsfxsize=3.5 in\centerline{\epsffile{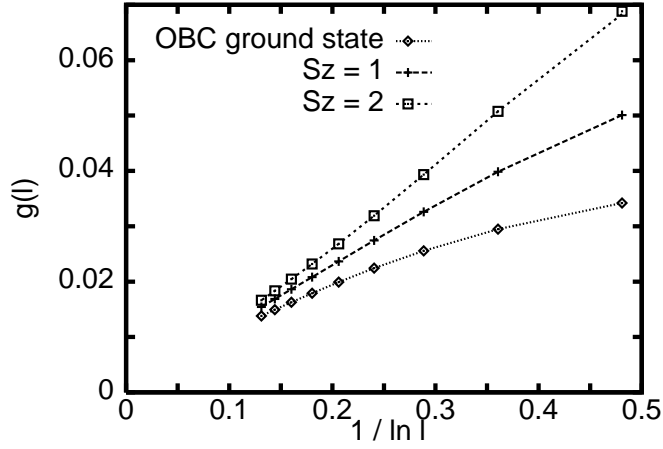}}
\vspace{0.5cm}
\caption[]{
$g(l)$ calculated by Eq.(\ref{oglz}) and Eq.(\ref{oglgs}) from Bethe 
ansatz energies of spin 1/2 antiferromagnetic Heisenberg open chain 
(OBC).  The three $g(l)$ are obtained from energies of ground state, 
total spin $S=1$ excited state and $S=2$ excited state, respectively.  
}
\label{figobcgl}
\end{figure}

\begin{figure}[ht]
\epsfxsize=3.5 in\centerline{\epsffile{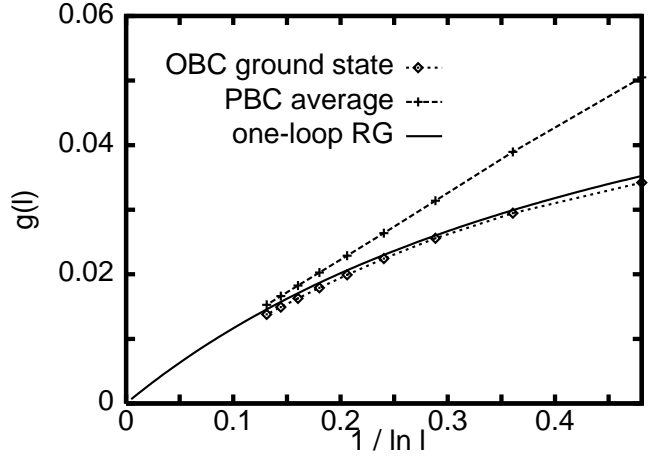}}
\vspace{0.5cm}
\caption[]{
The $g(l)$ calculated from ground state of open chain (OBC) and the 
average $g(l)$ for periodic\cite{Sorensen} chain (PBC), for spin 1/2 
antiferromagnetic Heisenberg model.  The full line is the one-loop 
renormalization group (RG) prediction\cite{Affleck1},
$g(l)=g_0(l_0)/[1+\pi b g_0(l_0) \ln(l/l_0)]$ with $b=4/\sqrt{3}$.  $g_0(l_0)$ is 
determined by the average of open chains ground state's $g(l)$ and 
the periodic chain's average $g(l)$ at the chain length $l_0=2048$.
}
\label{figavgl}
\end{figure}

To test the CFT predictions, we extract three estimates of $g(l)$ using
Eq.(\ref{oglz}) for the $S=1$ and $S=2$ excited states and Eq.(\ref{oglgs})
for the groundstate for chains up to two thousands sites by solving Eq.~(\ref{bae}) 
numerically.  We draw the three $g(l)$ completely determined by the 
energies of these three states, respectively, in 
Fig. \ref{figobcgl}. The reason that these three estimates of $g(l)$
don't agree exactly is because of the various corrections to these energies
of higher order in $g(l)$.  However, at large $l$ these estimates should
converge since $g(l)\to 0$.  
 We see that the coupling constants indeed collapse 
into one value which approaches $0$, 
as the chain length increases thus verifying the 
CFT predictions.  Previous numerical studies for periodic 
boundary conditions have verified finite size scaling obtained by conformal field 
theory\cite{Affleck1,Ng,Sorensen}.  We then compare the $g(l)$ obtained 
from the open boundary conditions with the one from periodic boundary 
conditions in Fig. \ref{figavgl}.  We redraw the ground state $g(l)$ 
for open boundary conditions and the average $g(l)$ given by the ground 
state, the singlet excitation, and the triplet excitation for periodic 
boundary conditions.  These data for periodic boundary conditions were 
obtained in Ref. [\onlinecite{Sorensen}].  The two $g(l)$ approach 
each other in the large length limit.  The one-loop 
renormalization group prediction for $g(l)$ given in 
Ref. [\onlinecite{Affleck1}], 
$g(l)=g_0(l_0)/[1+\pi b g_0(l_0) \ln(l/l_0)]$, with $b=4/\sqrt{3}$ is also drawn in 
Fig. \ref{figavgl}.  We use the average of the $g(l)$ for periodic and 
open chains at $l_0=2048$ to fix $g_0$.  The one-loop renormalization 
group prediction fits $g(l)$ at large $l$.  So we see that the 
logarithmic corrections for Heisenberg chains have been successfully 
predicted by conformal field theory.

\section{Correlation Functions and $1/T_1$}
Another application of the anomalous dimension of boundary operators is 
to Green's functions for a semi-infinite system.  A time-dependent Green's
function at the boundary, for  obeys the RG equation:
\begin{equation}
[\partial /\partial \ln \tau + \beta (g) \partial /\partial g + 2\gamma (g)]G(\tau ,g)=0,
\label{RGG}\end{equation}
where $\gamma$ is the anomalous dimension of the boundary operator whose
Green's function is being calculated.  This is given, to O(g), by Eq. (\ref{anom}).  $g(\tau )$
in Eq. (\ref{RGG}) is the effective coupling constant at scale $\tau$.
Solving this equation we obtain:
\begin{equation}
<\phi (\tau ,0)\phi (0,0)> \to 
	{(\ln |\tau |)^{-4b_n/b}\over \tau^{2x_n}}.
\end{equation}
Both the exponent, $x_n$ and the power of the logarithm are different than what occurs in the bulk.  
For the lowest dimension boundary operator of spin $S$ the factor of 
$\ln |\tau |$ is raised to the power $S_L(S_L+1)$.  In particular, for 
the spin operator at the boundary in the lattice Heisenberg model, the 
correlation function behaves as:
\begin{equation}
<\vec S_0(\tau )\cdot \vec S_0(0)> \to 
	\hbox{constant}{(\ln |\tau |)^2\over |\tau |^2}.
\label{corr}
\end{equation}
The imaginary part of the retarded Green's function at zero frequency 
and finite $T$, obtained from the Fourier transform of 
Eq. (\ref{corr}), gives the nuclear magnetic resonance relaxation rate, 
$1/T_1$ for a chain with non-magnetic impurities.  This behaves as:
\begin{equation}
1/T_1\propto T[\ln (T_0/T)]^2,
\end{equation}
for some temperature scale $T_0$ of order the exchange energy.  The 
$1/\tau^2$ power law was first derived in [\onlinecite{Eggert2}], 
without consideration of logarithmic corrections.  The discussion of 
the linear power-law in $1/T_1$, resulting from performing the Fourier 
transform, was first discussed, as far as we know, in 
[\onlinecite{Brunel}].  These authors also attempted to calculate the 
logarithmic correction.  However, their result, for which no derivation 
was given, differs from ours, containing $[\ln (T_0/T)]^4$ rather than 
$[\ln T_0/T)]^2$.  This behaviour is to be contrasted with that for the 
pure system, in which $1/T_1\propto [\ln (T/T_0)]^{1/2}$, constant up 
to a log correction.\cite{Sachdev}  

Of course, an actual experiment on a doped S=1/2 chain compound would
presumably average over all  distances from the chain ends. (This is 
related both to the fact that the relaxing nuclei can be at arbitrary 
locations and that even a nucleus near the end of a chain will have a 
transferred hyperfine interaction with spins further away from the 
chain end.)  We note that, at $T=0$ and ignoring log corrections, the 
spin self-correlation for a spin a distance $x$ from the chain end is
given by\cite{Eggert2}:
\begin{equation}
<\vec S_j(t)\cdot \vec S_j(0)> \propto 
{2x/v\over |t|\sqrt{t^2-4x^2/v^2}},
\end{equation}
where $v$ is the spin-wave velocity.  At sufficiently long times this 
decays as $1/t^2$ for all $x$.  However, for $t<<x/v$ it exhibits the 
bulk behaviour, decaying as $1/t$.  Thus we expect that the zero 
frequency finite T Fourier transform, which determines $1/T_1$, will be 
essentially constant (ignoring log corrections) down to a temperature
of order $v/x$, below which it will vanish essentially linearly in $T$. 

$1/T_1$ has been measured\cite{Takigawa2} for the quasi one dimensional antiferromagnetic
compound Sr$_2$CuO$_3$ obtaining apparent agreement with the
field theory prediction,\cite{Sachdev} $(\ln T)^{1/2}$.  Broad shoulders observed in the
NMR intensity\cite{Takigawa3} were interpreted as resulting from the distribution of 
local susceptibilities predicted by field theory methods\cite{Eggert3} for
chains with free ends.  Possibly such data will also verify the distribution
of relaxation rates resulting from impurities.  \\ \\ 

After this work was completed we managed to obtain a copy of a three year old preprint,\cite{Kawano}
which was never published nor available on the xxx archive, and which derived many of the results obtained here.
\acknowledgements
We would like to thank C. Itoi for helpful discussions and H. Asakawa for sending us a copy
of Ref. (\onlinecite{Kawano}).  This research was supported in part by NSERC of Canada.


\begin{references}
\bibitem{Cardy} J.L. Cardy, J. Phys. {\bf A19}, L1093 (1986).
\bibitem{Affleck1} 
I. Affleck, D. Gepner, H.J. Schulz and T. Ziman, 
J. Phys. {\bf A22}, 511 (1989).
\bibitem{Singh} R.R.P. Singh, M.E. Fisher and R. Shankar, Phys. Rev. {\bf B39}, 2562 (1989).
\bibitem{Eggert1} S. Eggert, I. Affleck and M. Takahashi, Phys. Rev. Lett. {\bf 73}, 323 (1994).
\bibitem{Sachdev} S. Sachdev, Phys. Rev. {\bf B50}, 13006 (1994).
\bibitem{Motoyama} N. Motoyama, H. Eisaki and S. Uchida, Phys. Rev. Lett. {\bf 76}, 3212 (1996).
\bibitem{Takigawa1} M. Takigawa, O.A. Starykh, A.W. Sandvik and R.R.P. Singh, Phys. Rev. {\bf B56}, 13681 (1997).
\bibitem{Eggert2}
S. Eggert and I. Affleck, Phys. Rev. {\bf B46}, 10866 (1992).
\bibitem{Affleck2} For a review see I. Affleck, Acta Physica Polonica, {\bf 26}, 1869 (1995); cond-mat/9512099.
\bibitem{Ng} 
T. K. Ng, S. Qin and Z. B. Su, Phys. Rev. {\bf B54}, 9854 (1996).
\bibitem{Blote}
H.W.J. Bl\"ote, J.L. Cardy and M.P. Nightingale, 
Phys. Rev. Lett. {\bf 56}, 742 (1986).
\bibitem{Hamer} C.J. Hamer, G.R.W. Qispel and M.T. Batchelor, J. Phys. {\bf A20}, 5677 (1987).
\bibitem{Gaudin} 
M. Gaudin, Phys. Rev. {\bf A4}, 386 (1971).
\bibitem{Alcarez} F.C. Alcarez, M.N. Barber, M.T. Batchelor, R.J. Baxter
and G.R.W. Quispel, J. Phys. {\bf A20}, 6397 (1987).
\bibitem{Batchelor} M.T. Batchelor and C.J. Hamer, J. Phys. {\bf A23}, 761 (1990).
\bibitem{Sorensen} 
E. S. S\o rensen, S. Eggert, and I. Affleck,
J. Phys. {\bf A26}, 6757 (1993).
\bibitem{Faddeev} 
L. Faddeev and L. A. Takhtajan, Phys. Lett. {\bf A85}, 375 (1981).
\bibitem{Brunel} 
V. Brunel, M. Bocquet and Th. Jolicoeur, preprint, cond-mat/9902028.
\bibitem{Takigawa2} M. Takigawa, N. Motoyama, H. Eisake and S. Uchida,
Phys. Rev. Lett. {\bf 76}, 4612 (1996).
\bibitem{Eggert3} S. Eggert amd I. Affleck, Phys. Rev. Lett. {\bf 75}, 934 (1995).
\bibitem{Takigawa3}  M. Takigawa, N. Motoyama, H. Eisake and S. Uchida,
Phys. Rev. {\bf B55}, 14129 (1997).
\bibitem{Kawano} K. Kawano and M. Yajima, unpublished.
\end{references}
\end{document}